\documentclass[%
reprint,
nofootinbib,
 amsmath,amssymb,
 aps,
]{revtex4-1}

\usepackage{graphicx}
\usepackage{dcolumn}
\usepackage{bm}
\usepackage[mathlines]{lineno}

\begin{document}


\title{Implications of gamma-ray observations on proton models of UHECR}

\author{A.D.~Supanitsky}
\affiliation{Instituto de Astronom\'ia y F\'isica del Espacio (IAFE, CONICET-UBA), CC 67, Suc.~28, C1428ZAA Buenos Aires, Argentina.}
\email{supanitsky@iafe.uba.ar}

\date{\today}

\begin{abstract}

The origin of ultra high energy cosmic rays (UHECR) is still unknown. However, great progress has been achieved in past 
years due to the good quality and large statistics in experimental data collected by the current observatories. The data of the 
Pierre Auger Observatory show that the composition of the UHECRs becomes progressively lighter starting from $10^{17}$ eV up 
to $\sim 10^{18.3}$ eV and then, beyond that energy, it becomes increasingly heavier. These analyses are subject to important 
systematic uncertainties due to the use of hadronic interaction models that extrapolate lower energy accelerator data to the 
highest energies. Although proton models of UHECRs are disfavored by these results, they cannot be completely ruled out. 
It is well known that the energy spectra of gamma rays and neutrinos, produced during propagation of these very energetic 
particles through the intergalactic medium, are a useful tool to constrain the spectrum models. In particular, it has 
recently been shown that the neutrino upper limits obtained by IceCube challenge the proton models at 95\% CL. In this 
work we study the constraints imposed by the extragalactic gamma-ray background, measured by Fermi-LAT, on proton models of 
UHECRs. In particular, we make use of the extragalactic gamma-ray background flux, integrated from 50 GeV to 2 TeV, that 
originates in point sources, which has recently been obtained by the Fermi-LAT collaboration, in combination with the neutrino 
upper limits, to constrain the emission of UHECRs at high redshits ($z>1$), in the context of the proton models.      
 
\end{abstract}

\maketitle


\allowdisplaybreaks 

\section{Introduction}

Despite a big experimental effort done in the past years, the origin of the ultra high energy cosmic rays 
(UHECRs), i.e. with energies larger than $10^{18}$ eV, is still unknown. The Pierre Auger \cite{Auger:15} and 
Telescope Array \cite{TA:12} observatories, are currently taking data in this energy range. The Southern 
hemisphere Pierre Auger Observatory, in the Province of Mendoza, Argentina, is the largest cosmic ray observatory 
in the world. The Telescope Array observatory, placed in Utah, USA, is the largest in the northern hemisphere. 
The UHECR energy spectrum presents two main features, the ankle, found at an energy 
of $\sim10^{18.7}$ eV, which consists in a hardening of the flux, and a suppression at $10^{19.6}-10^{19.7}$ eV 
\cite{Valino:15,Jui:15}. Note that, even though the spectra observed by The Pierre Auger and Telescope Array 
observatories present some differences, these two features are observed in both measurements.        
     
UHECRs generate atmospheric air showers when they interact with the molecules of the atmosphere. These air
showers can be observed by using ground detectors and/or fluorescence telescopes. The ground detectors detect 
the secondary particles, produced during the shower development, that reach the ground. Whereas, the fluorescence 
telescopes detect the fluorescence light emitted by the interaction of the secondary charged particles with the 
molecules of the atmosphere. The Pierre Auger and Telescope Array observatories have both types of detectors, and 
a subsample of the events are observed by both detector types. The arrival direction, the energy, and the composition 
of the primary UHECR have to be inferred from the air shower observations. In particular, the composition is 
obtained by comparing observables sensitive to the primary mass with simulations of the atmospheric air showers. 
This is subject to large systematic uncertainties because the hadronic interactions at the highest energies are 
unknown. There are models that extrapolate the lower energy accelerator data to the highest energies. Some of 
these models have been recently updated with the Large Hadron Collider data. As a consequence, the differences 
among the predictions of the different models decreased but did not disappear. 

One of the parameters most sensitive to primary mass is the atmospheric depth of the shower maximum, $X_{max}$. 
This parameter can be reconstructed with the data taken by fluorescence telescopes. The composition analyses 
based on $X_{max}$ observed by Auger, done by using the updated hadronic interaction models, show a decrease 
of the primary mass from $10^{17}$ eV up to $\sim 10^{18.3}$ eV and then, beyond that energy, the mass becomes 
increasingly heavy \cite{Porcell:15,AugerXmax:14}. On the other hand, the $X_{max}$ data obtained by Telescope 
Array is consistent with protons \cite{TAXmax:15,Fujii:15}. However, it has been recently shown that the 
composition data observed by the two experiments are consistent \cite{Abbasi:15}. It is worth mentioning that, 
the statistics of Telescope Array is smaller than the one corresponding to Auger.

The interpretation of the UHECR spectrum depends strongly on the mass composition. If protons are the dominant
component, the ankle is originated by pair production in the interaction of the protons with the low energy 
photons of the extragalactic background light (EBL) and cosmic microwave background (CMB), during propagation 
through the intergalactic medium. The suppression observed at the highest energies can originate by the 
photopion-production of protons interacting with the CMB photons, by the intrinsic cutoff of in the sources, 
or by a combination of both effects. The proton model (known as the dip model) of the UHECR spectrum have been 
extensively studied in the literature (see 
Refs.~\cite{Hill:85,DeMarco:03,Bere:04,Bere:05,Bere:06,Aloisio:07,Aloisio:08}). 

On the other hand, if the UHECR spectrum is composed by heavy nuclei besides protons, the ankle can be interpreted 
as the transition between the galactic and extragalactic cosmic rays (see Ref.~\cite{AloisioRev:12} for a review). 
However, this possibility is disfavoured by the Auger data \cite{Auger:12}, large scale anisotropy studies show 
that the extragalactic component should continue below the ankle. Recently, a new light extragalactic component,
that dominates the flux below the ankle, originated in different sources than the ones responsible of 
the flux at the highest energies, has been proposed \cite{Aloisio:14}. This low energy component could also 
originate in the photodisintegration of nuclei with the photon fields in the vicinity of the acceleration 
region or the source \cite{Unger:15,Allard:15}. In these scenarios the suppression is due to the 
photodisintegration process undergone by the nuclei when they interact with photons of the EBL and CMB, to an 
intrinsic spectral cutoff in the injected spectrum, or to a combination of both effects.   
 
Besides composition information, the observation of secondary gamma rays and neutrinos, generated in the interactions 
undergone by the cosmic rays during propagation through the universe, can constrain the different models of the UHECR 
spectrum. In particular, a smaller number of gamma rays and neutrinos are predicted in scenarios with larger fractions 
of heavy nuclei at the highest energies. In Ref.~\cite{Heinze:15} the dip model has been rejected at 95\% CL by using 
the upper limit on the neutrino spectrum obtained by the IceCube experiment \cite{Ishihara:15}. In that work the source 
evolution is assumed to be the one corresponding to the star formation rate times $(1+z)^m$ where $z$ is the redshift 
and $m$ is a parameter fixed by fitting Telescope Array data. However, it is also shown that models with no emission 
at $z>1$ cannot be rejected with present neutrino data. 
         
Extragalactic gamma-ray background (EGB) observations impose quite restrictive constrains on models of the UHECR 
spectrum \cite{Bere:75,Fodor:03,Semikoz:04,Ahlers:10,Decerprit:11,Bere:11,Hooper:11,Gelmini:12}. The EGB has recently 
been measured, from 100 MeV to 820 GeV, by Fermi-LAT \cite{Ackermann:15}. Part of the EGB originates in gamma-ray point 
sources. The gamma rays generated by the propagation of UHECRs can contribute to a diffuse component. By using the 
2FHL catalog \cite{Ackermann:16} of 360 point sources (mostly blazars) detected by Fermi-LAT, it has been shown that 
$86^{+16}_{-14}\%$ of the integrated EGB spectrum from 50 GeV to 2 TeV originates in point sources \cite{AckermannPS:16}. 
By using this result, in Ref.~\cite{Liu:16} it is found that only a group of nearby sources, contributing in the energy 
range below the ankle, can be responsible for the light component observed in that energy range. Also, in 
Ref.~\cite{Kalashev:16}, a more restrictive upper limit to the energy density of the electromagnetic cascades, that are
developed in the intergalactic medium, is found by using this new result.     
  
In this work we study the impact of this new estimation on proton models of UHECRs. Instead of using the upper limit to the 
energy density found in Ref.~\cite{Kalashev:16}, we calculate an upper limit to the integrated gamma-ray flux, in the 
energy range from 50 GeV to 2 TeV, that does not originate in point sources. This upper limit is used to study the 
constraints that this new result imposes on proton models of the UHECRs spectrum, assuming a more relaxed source evolution 
than the one used in Ref.~\cite{Heinze:15}. We also compare the results obtained by using the gamma-ray information with 
the ones obtained by using the neutrino upper limit.

\section{Fit of the UHECR spectrum}
\label{Main}

The Telescope Array measurement is considered to fit the UHECR flux assuming a pure proton component. This is 
because, as mentioned above, the Telescope Array data are compatible with a proton composition at the highest 
energies. Following Refs.~\cite{Kido:15,Heinze:15} the fits are performed for energies above $10^{18.2}$ eV.

It is assumed that the sources are uniformly distributed in the Universe and that the injected spectrum follows a power 
law with an exponential cutoff,
\begin{equation}
\phi(E,z) = C\ S(z)\ E^{-\alpha} \exp(-E/E_{cut}),
\label{Jinj}
\end{equation}   
where $C$ is a normalization constant, $\alpha$ is the spectral index, $E_{cut}$ is the cutoff energy, and $S(z)$ parametrize
the source density and luminosity evolution of the injection with redshift $z$. The evolution function $S(z)$ is not known,
it depends on the types of sources responsible of the acceleration of the cosmic rays, which are still unknown. The most common
options found in the literature correspond to the evolution of the star formation rate (SFR), the gamma-ray bursts (GRB), and 
the active galactic nuclei (AGN) (see for instance Refs.~\cite{Heinze:15,Gelmini:12,Kalashev:13}). In this work a broken power 
law of $(1+z)$ is assumed,
\begin{equation}
S(z) = \left\{ 
\begin{array}{ll}
(1+z)^m & z\leq 1 \\
2^{m-n}\ (1+z)^n & z>1\ \&\ z\leq 6 \\
%
0 & z>6 
\end{array}    \right., 
\end{equation}
where $m$ and $n$ are free parameters. The break point at $z=1$ is motivated by the SFR, GRB, and some special cases of AGN
evolution functions \cite{Kalashev:13}. Note that just the parameter $m$ is important for the fit of the spectrum because the 
contribution of the sources at $z>1$ is negligible for $E \geq 10^{18.2}$ eV. The evolution corresponding to $z>1$ is important 
for the generation of gamma rays and neutrinos.

The energy spectrum of UHECR at Earth is calculated by using the program TransportCR \cite{Kalashev:14}, which has been
developed to solve numerically the transport equations that governs the propagation of the cosmic rays in the intergalactic medium.
One of the advantages of this approach is the decrease of the computational time compared with the one corresponding to the methods
based on the Monte Carlo technique. This makes it very useful for fitting the cosmic ray energy spectrum. Besides nuclei, the 
TransportCR code propagates gamma rays and neutrinos produced in the interactions of them with the photon backgrounds. The 
photopion production is based on the SOPHIA code \cite{sophia} and there are several models of the EBL available.

Models with fixed cutoff energy are considered. It is possible to fit also $E_{cut}$ as it is done in Ref.~\cite{Heinze:15}, but
the computational time increases considerably. Note that the processing time is larger in the case considered in this work because 
of the calculation of the propagation of gamma rays and neutrinos. For a given cutoff energy the proton flux at Earth is calculated
for a discrete grid of the parameters $\alpha$, $m$, and $n$. As mentioned before, the fit of the spectrum depends on the parameters 
$\alpha$ and $m$ in the energy range under consideration, then the parameter $n$ just modifies the flux of secondary particles
without affecting the fit. That is why only some particular values of $n$ are considered. The spectrum for any value of energy,
$\alpha$, and $m$ is interpolated by using a trilinear interpolation algorithm, for fixed values of $n$.

The fit of the Telescope Array spectrum is performed by minimizing the chi-square, which is given by,
\begin{equation}
\chi^2=\sum_{i=1}^N \frac{(J(\tilde{E}_i,\alpha,m,C)-J_i)^2}{\sigma_i^2} + \left( \frac{\delta_E}{\sigma_E} \right)^2
\label{chi2}
\end{equation}                    
where $J(E,\alpha,m,C)$ is the interpolated spectrum, $\tilde{E}_i=(1+\delta_E) E_i$ with $E_i$ the energy of the $i$-th data point
and $\delta_E$ is a systematic shift in the energy scale (see Ref.~\cite{Heinze:15,Kido:15}), $J_i$ is the flux measured corresponding 
to the $i$-th bin, and $\sigma_i^2$ is the uncertainty in the determination of the flux also for the $i$-th bin. The last term in the 
$\chi^2$ takes into account the systematic uncertainty on the determination of the energy, where $\sigma_E=0.2$ (see 
Ref.~\cite{Heinze:15}). Following Ref.~\cite{Heinze:15} $C$ and $\delta_E$ are considered as nuisance parameters and then the profile 
likelihood technique \cite{Agashe:14} is used to get rid of of them. In this technique the nuisance parameters are replaced by their 
maximum likelihood estimators. Therefore, only $m$ and $\alpha$ are the parameters of interest.      
 
The EBL is still quite uncertain, principally because a direct measurement is subject to several foregrounds \cite{Cooray:16}. 
However, there are different models, available in the literature, for the spectrum as a function of the photon energy and redshift. 
In Ref.~\cite{Kido:15} it was shown that the fit of the cosmic ray energy spectrum, assuming a pure proton composition, depends on 
the EBL model considered. Moreover, the gamma-ray and neutrino spectra also depend on the EBL model assumed (see 
Refs.~\cite{Kalashev:16} and \cite{Stanev:05} for gamma rays and neutrinos, respectively). In this work we consider the EBL model 
of Ref.~\cite{Kneiske:04}, which is commonly used to model the UHECR flux and also, following Ref.~\cite{Kalashev:16}, the one of 
Ref.~\cite{Inoue:12}.   
 
\begin{figure}[!ht]
\includegraphics[width=8cm]{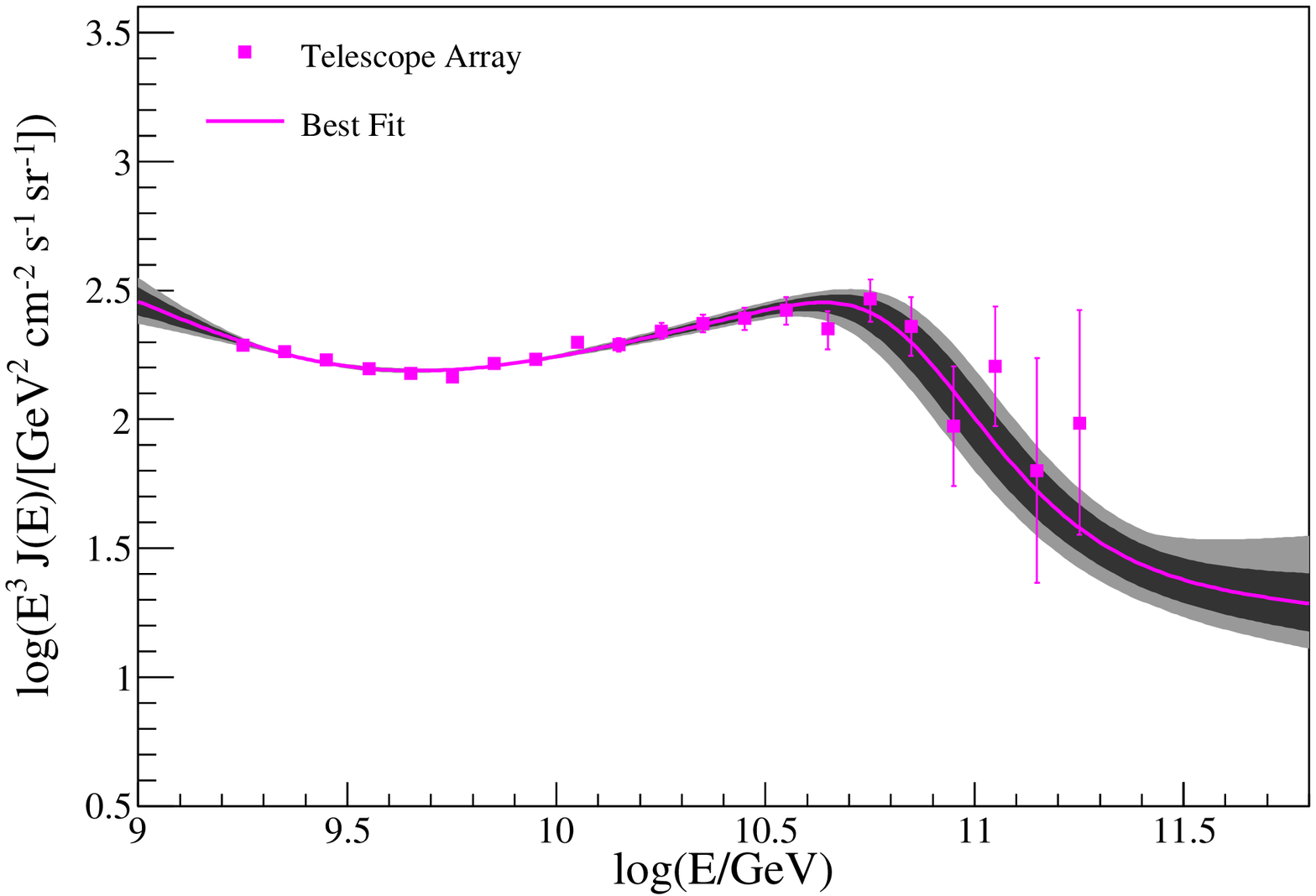}
\includegraphics[width=8cm]{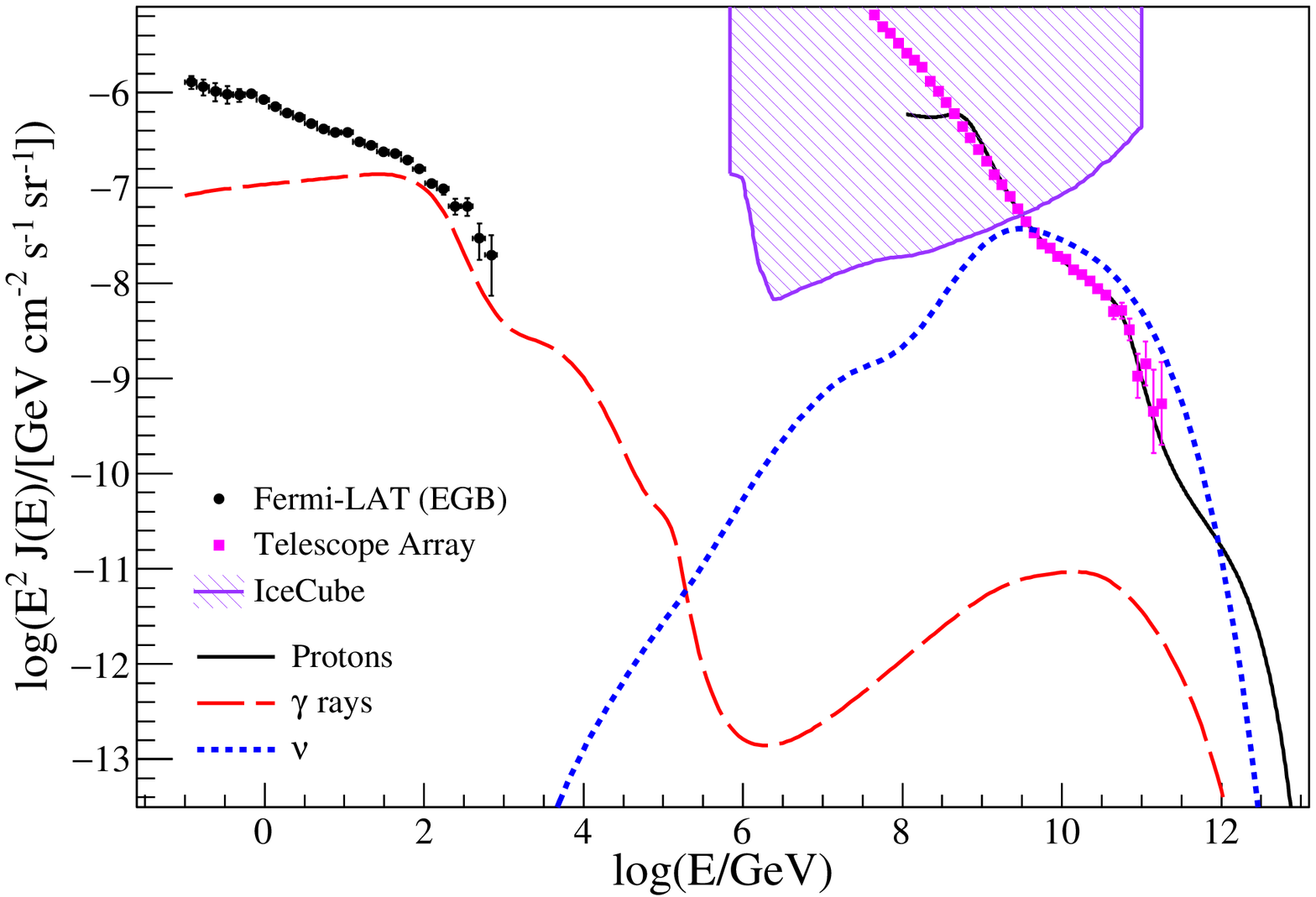}
\caption{Top panel: Best fit of the Telescope Array energy spectrum for the EBL model of Ref.~\cite{Kneiske:04}, no emission above $z=1$ 
($n \rightarrow -\infty$), and for  $E_{cut}=10^{21}$ eV. The shadowed areas correspond to 68.27\% and 95.45\% CL regions. Bottom panel: proton, 
gamma-ray, and neutrino spectra corresponding to the best fit of the Telescope Array data. The filled circles correspond to the EGB obtained 
by Fermi-LAT and the shadowed area corresponds to the rejection region at 90\% CL for the upper limit on the neutrino flux obtained by IceCube. 
\label{KN21Inf}}
\end{figure}

The top panel of Fig.~\ref{KN21Inf} shows the fit of the Telescope Array energy spectrum considering the EBL model of 
Ref.~\cite{Kneiske:04}, no cosmic ray emission above $z=1$, i.e. $n \rightarrow -\infty$, and $E_{cut}=10^{21}$ eV. The 
Telescope Array spectrum is taken from \cite{Kido:15} and correspond to the spectrum obtained by the surface 
detectors\footnote{Instead of the combined spectrum, reported in Ref.~\cite{Jui:15}, the one obtained by the surface detectors, 
reported and fitted in Ref.~\cite{Kido:15}, is considered for the fit because the error bars at low energies can be extracted 
from the plot in that paper, which is not the case for the plot in Ref.~\cite{Jui:15}.}. The energy scale of the proton spectrum 
has been shifted according to the value $\delta_E$ obtained. The best fit is obtained for $\alpha = 2.16$, $m = 6.78$, and 
$\delta_E = -0.096$. Note that this results are compatible with the ones obtained in Ref.~\cite{Kido:15} even though 
they do not use the penalty term in the $\chi^2$. The bottom panel of Fig.~\ref{KN21Inf} shows the proton, gamma-ray, and 
neutrino spectra corresponding to the best fit. Also in this case, the energy scale of the proton spectrum has been shifted 
according to the value $\delta_E$ obtained. The plot also shows the upper limit on the neutrino flux obtained by IceCube at 
90\% CL \cite{Ishihara:15}. The data points at low energies correspond to the spectrum of the EGB observed by Fermi-LAT
\cite{Ackermann:15}. Note that the best fit is compatible with the neutrino upper limit an also with the EGB. It is worth 
mentioning that the Telescope Array energy spectrum is extended to energies below $10^{18.2}$ eV with the combined spectrum 
taken from Ref.~\cite{Jui:15}. It can also be seen that the fit overshoots the data points below $10^{18.2}$ eV. This low 
energy contribution can be suppressed due to the diffusion of the particles in the intergalactic magnetic field 
\cite{Berezinsky:07} or by the presence of a low energy cutoff in the injection spectrum (see Ref.~\cite{Heinze:15} for 
details).         

As mentioned in the introduction, the EGB measurement has important implications on proton models of the UHECR spectrum. By 
using the results obtained in Ref.~\cite{AckermannPS:16}, an upper limit on the integrated flux corresponding to the component of the 
EGB spectrum that do not originate in point sources is obtained,
\begin{equation}
I_\gamma^{UL} = 9.354 \times 10^{-10} \textrm{cm}^{-2}\ \textrm{s}^{-1} \textrm{sr}^{-1}, 
\end{equation}
at 90\% CL. Here the energy range considered for the integral goes from 50 GeV up to 2 TeV. The method developed by Feldman \& Cousins 
\cite{Feldman:98} is used to obtain the upper limit (see Appendix \ref{UL} for the details of the calculation). 

For each pair $\alpha$ and $m$ considered to fit the Telescope Array data, after applying the profile likelihood method to the parameters $C$ 
and $\delta_E$ as described above, the integral of the gamma-ray energy spectrum is calculated and the region in the $\alpha-m$ plane, defined 
by,
\begin{equation}
\int_{50\ \textrm{GeV}}^{2\ \textrm{TeV}} dE\ J_\gamma(E,\alpha,m) \leq I_\gamma^{UL},
\label{Iulreg}
\end{equation}  
is determined.

\begin{figure}[!ht]
\includegraphics[width=8cm]{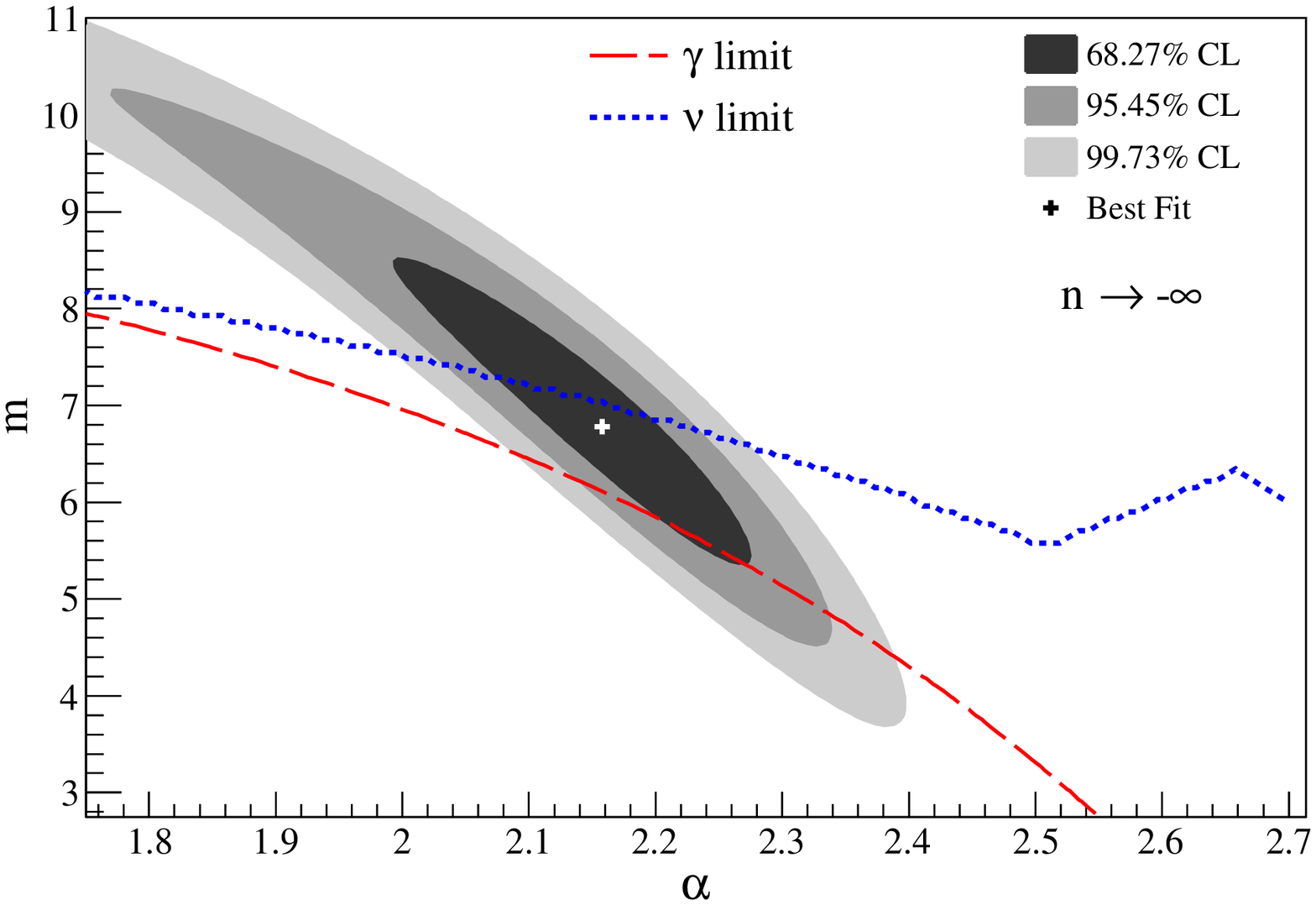}
\includegraphics[width=8cm]{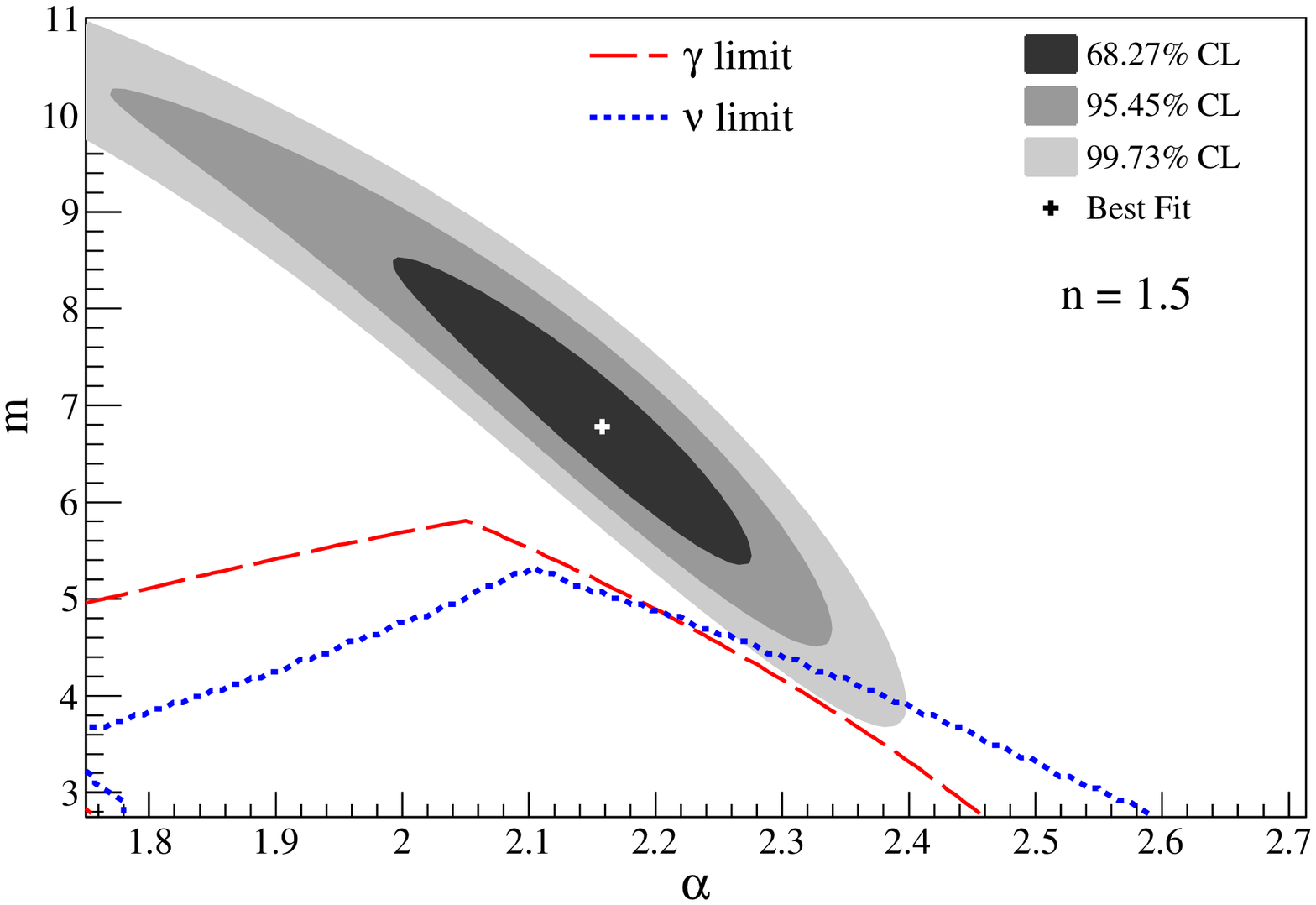}
\caption{Top panel: Best fit and confidence regions for $n \rightarrow -\infty$. The allowed regions corresponding to gamma-ray and neutrino 
observations are below the dashed and dotted curves, respectively. Bottom panel: Best fit and confidence regions for $n=1.5$. The allowed regions 
corresponding to gamma-ray and neutrino observations are below the larger dashed and dotted curves and above the smaller ones in the bottom-left 
corner, respectively. The EBL model of Ref.~\cite{Kneiske:04} is considered and $E_{cut}=10^{21}$ eV. \label{RegKN21}}
\end{figure}

The top panel of Fig.~\ref{RegKN21} shows the best fit and the regions of 68.27\%, 95.45\%, and 99.73\% CL for the fit in 
Fig.~\ref{KN21Inf}. Also shown are the allowed regions inferred from the upper limits obtained form Eq.~(\ref{Iulreg}) for 
the case of gamma rays and from the condition $J_\nu(E,m,n) \leq J_\nu^{UL}(E)$, where $J_\nu^{UL}(E)$ is the upper limit on 
the flux obtained by IceCube (see bottom panel of Fig.~\ref{KN21Inf}), for the case of neutrinos. It can be seen from the plot 
that, while the best fit is compatible with the neutrino data, it is in tension with the upper limit obtained from the non-point 
sources component of the EGB. Moreover, the best fit is also in tension with the upper limit obtained at 99\% CL (see Appendix 
\ref{99}). In this case the upper limit obtained from gamma-ray data is more restrictive than the one coming from the neutrino 
data. The bottom panel of the figure shows the result obtained in the same conditions as the ones corresponding to the top panel 
but for $n=1.5$. It can be seen that the best fit and the allowed regions are, as expected, unaltered, but in this case even 
the region corresponding to 99.73\% CL is in tension with the gamma-ray upper limit. Any value of the parameter $n$ larger than 
$1.5$ has associated a larger production of secondary gamma rays and neutrinos. Therefore, the region corresponding to 99.73\% CL 
of the fit is also in tension with at least the gamma-ray upper limit for any value of $n$ larger than 1.5.           

\begin{figure}[!ht]
\includegraphics[width=8cm]{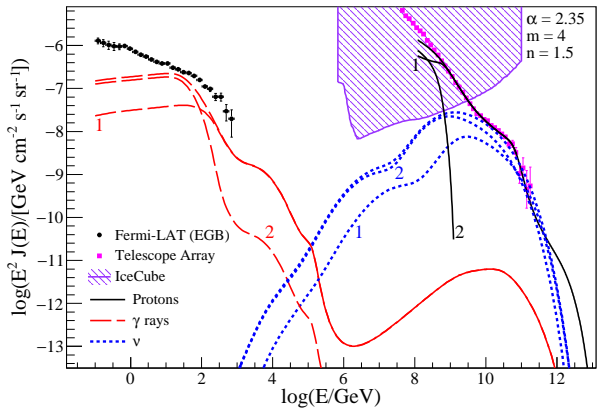}
\caption{Proton, gamma-ray and neutrino spectra for $\alpha=2.35$, $m=4$, and $n=1.5$. The curves labelled as 1 correspond
to sources at $z \in [0,1]$ and the ones labelled as 2 correspond to sources at $z \in [1,6]$. Also shown are the total 
spectra obtained by adding the components 1 and 2.\label{SpecZ12}}
\end{figure}

From Fig.~\ref{RegKN21} it can be seen that for the case in which the UHECRs are generated in the redshift range from $z=0$ to 
$z=1$, the gamma-ray upper limit imposes more restrictive conditions than the ones corresponding to the neutrino upper limit. 
However, when the sources produce UHECRs beyond $z=1$ the restrictions obtained from gamma-ray and neutrino observations are
complementary, i.e. by using both, the neutrino and gamma-ray upper limits, it is possible to enlarge the rejection region. The 
loss of restrictive power of the gamma-ray upper limit is due to the fact that the attenuation of gamma rays, owing to pair
production with the EBL photons, that originate in $z > 1$ is larger than the one corresponding to the gamma rays originated in 
$z \leqslant 1$. This can be seen from Fig.~\ref{SpecZ12} where the contributions from $z \in [0,1]$ and $z \in [1,6]$ to the 
total spectra are shown. The case considered corresponds to $\alpha=2.35$, $m=4$, and $n=1.5$, which is a point inside the 
99.73\% CL region of the fit for $n=1.5$. This point is also in the rejection region obtained from the gamma-ray data but in 
the allowed region obtained from the neutrino data. In fact while the neutrino flux increases by a factor of $\sim 5.5$ at 
$10^9$ GeV, when the component corresponding to $z \in [1,6]$ is added, the integral of the gamma-ray spectrum increases in a 
factor of $\sim 2.2$.  

\begin{figure}[!ht]
\includegraphics[width=8cm]{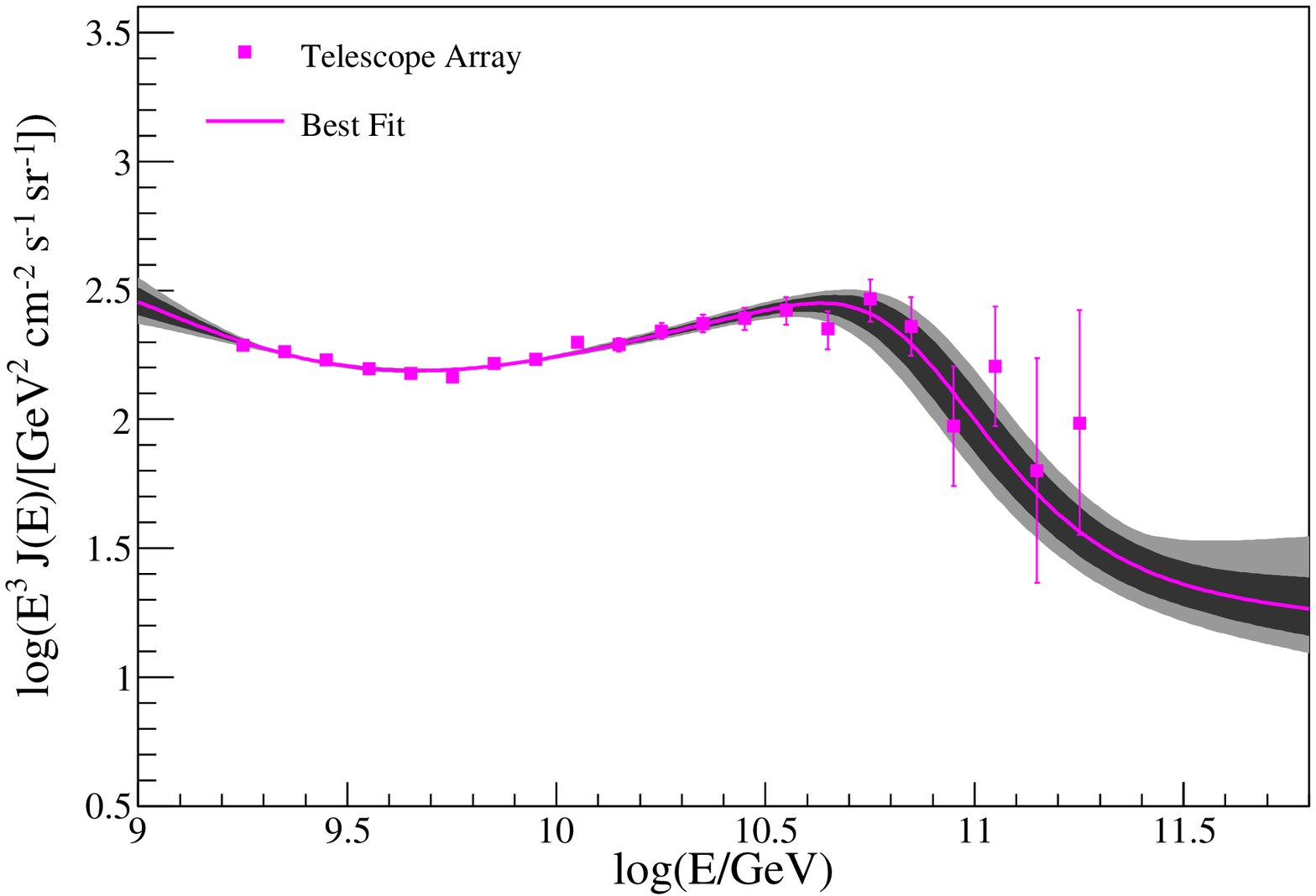}
\includegraphics[width=8cm]{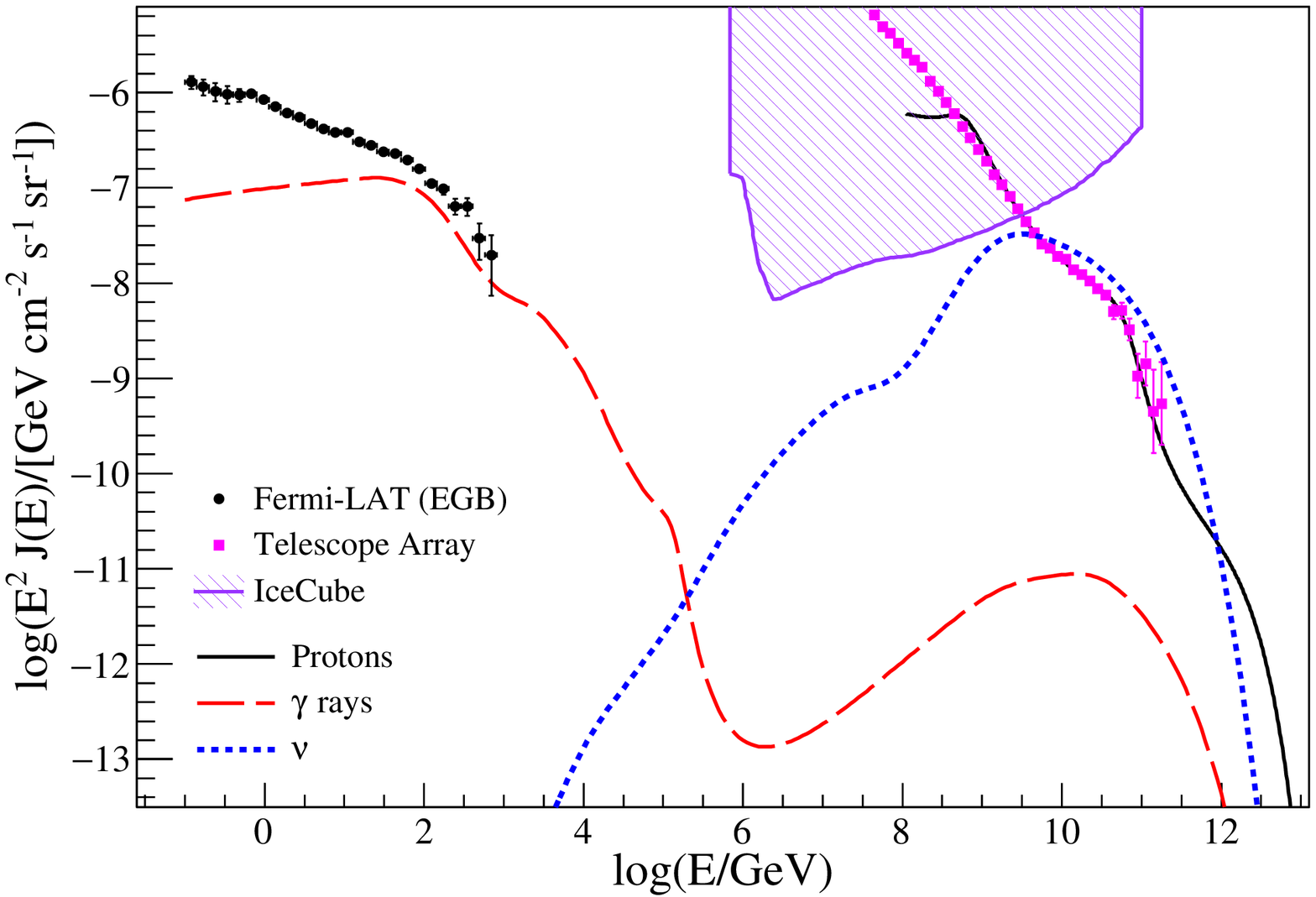}
\caption{Same as Fig.~\ref{KN21Inf} but for the EBL model of Ref.~\cite{Inoue:12}. \label{IN21Inf}}
\end{figure}

Figure \ref{IN21Inf} shows the best fit and the associated gamma-ray and neutrino spectra for the EBL model of Ref.~\cite{Inoue:12}, 
for no UHECR emission above $z=1$, and for $E_{cut}=10^{21}$ eV. The best fit is obtained for $\alpha = 2.17$, $m = 6.65$, and 
$\delta_E = -0.10$. Although, the best fit parameters are different from the ones obtained for the other EBL model considered 
before, the differences between Figs.~\ref{IN21Inf} and \ref{KN21Inf} are small. 

\begin{figure}[!ht]
\includegraphics[width=8cm]{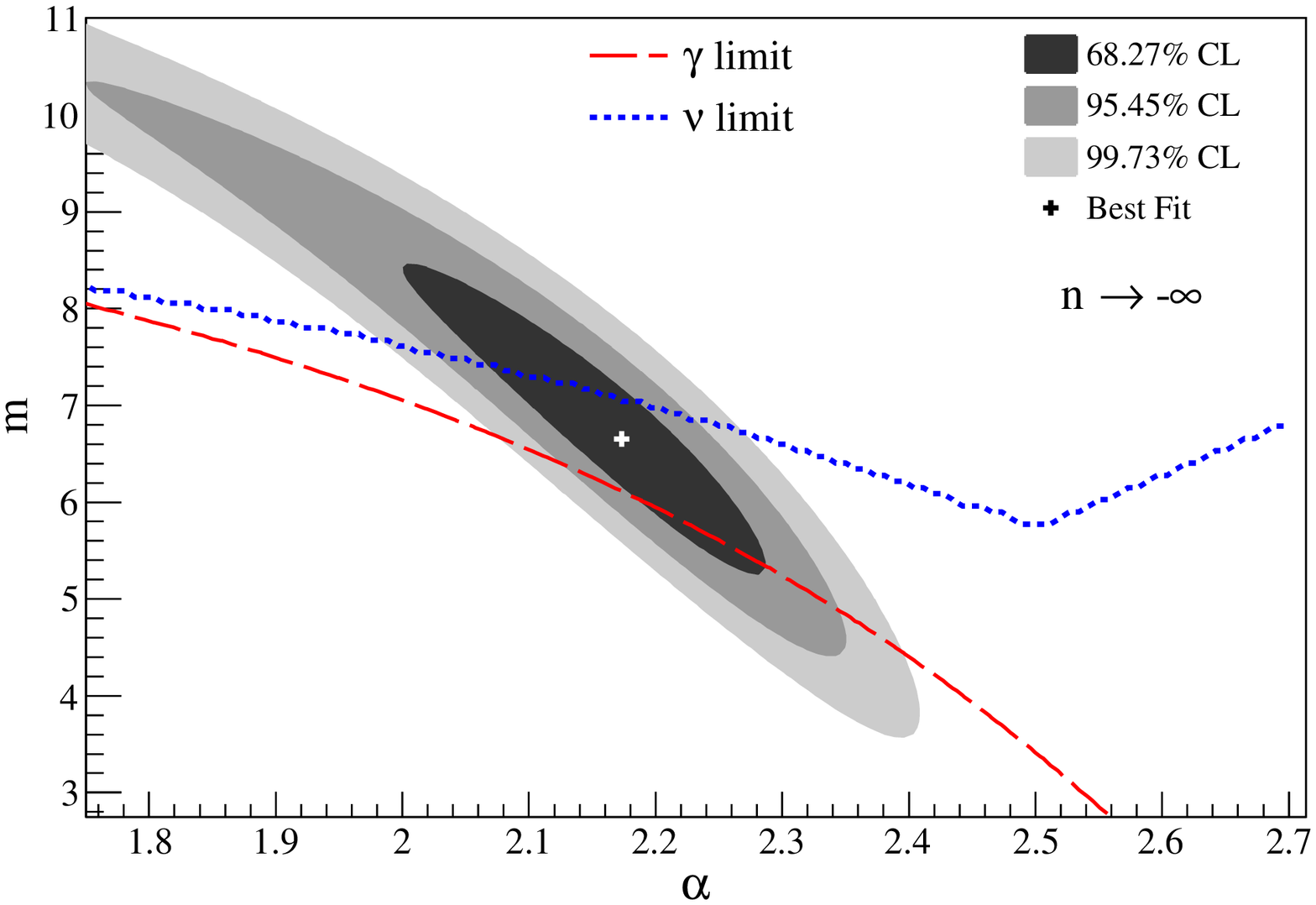}
\includegraphics[width=8cm]{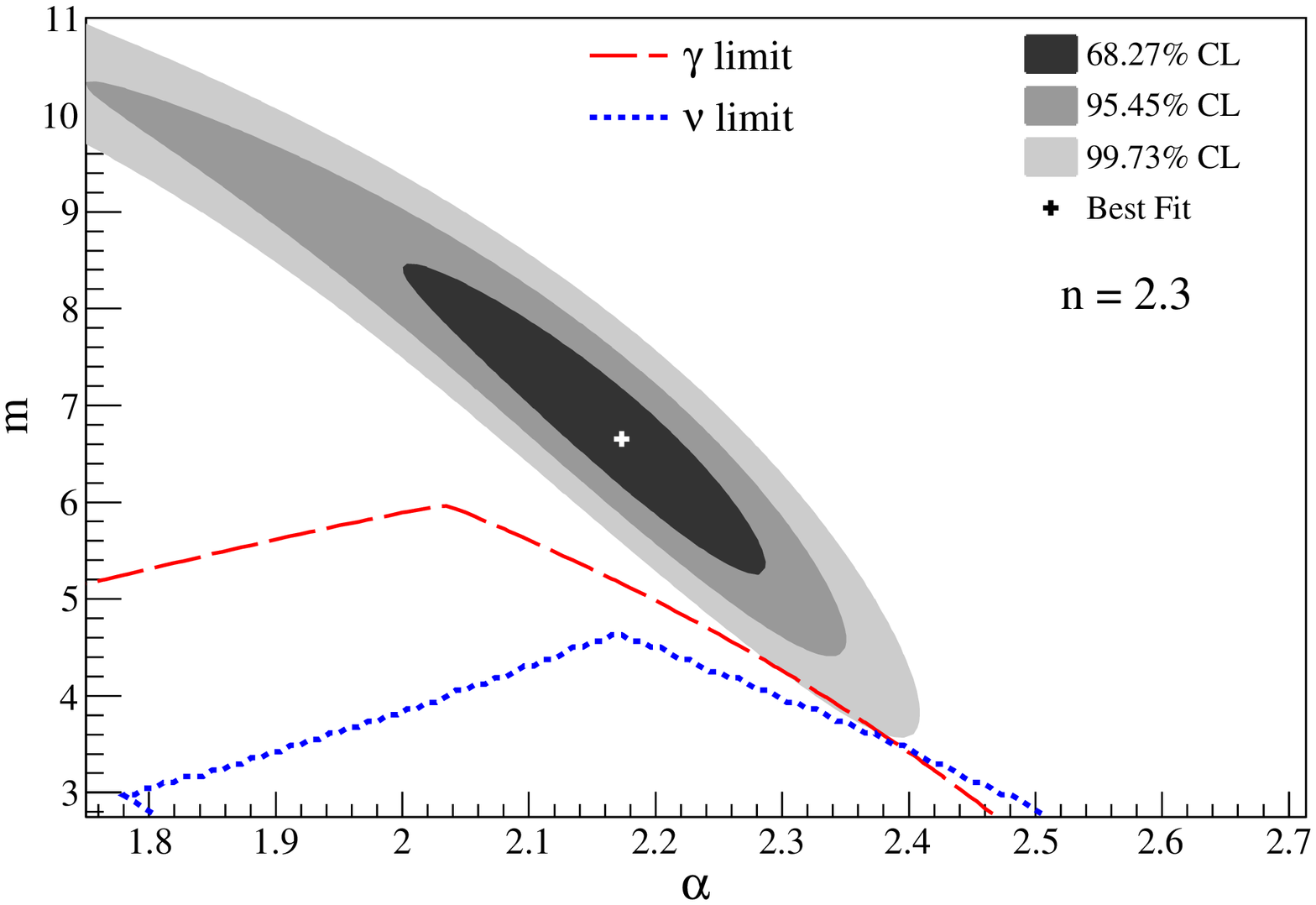}
\caption{Top panel: Best fit and confidence regions for $n \rightarrow -\infty$. The allowed regions corresponding to gamma-rays and neutrino 
observations are below the dashed and dotted curves, respectively. Bottom panel: Best fit and confidence regions for $n=2.3$. The allowed region 
from the gamma-ray upper limit is below the dashed curve. The allowed region from the neutrino upper limit is below the large dotted curve 
and above the small dotted curve in the bottom-left corner. The EBL model of Ref.~\cite{Inoue:12} is considered and $E_{cut}=10^{21}$ eV.
\label{RegIN21}}
\end{figure}

The top panel in Fig.~\ref{RegIN21} shows the best fit parameters and the regions of 68.27\%, 95.45\%, and 99.73\% CL for the 
fit of Fig.~\ref{IN21Inf}. Also shown are the allowed regions inferred from the gamma-ray and neutrino upper limits. Also in 
this case, the best fit is compatible with the neutrino upper limit but it is in tension with the gamma-ray upper limit at both 
90\% and 99\% CL (see Appendix \ref{99}). Comparing with the top panel of Fig.~\ref{RegKN21}, it can be seen that, in this case, 
the upper limits coming from gamma-ray and neutrino data are less restrictive. This is because the spectra of these secondary
particles, predicted by using the second EBL model considered, are smaller in almost the entire energy range relevant for each 
type of particle.     

The bottom panel of Fig.~\ref{RegIN21} shows the confidence regions of the fit and the allowed regions inferred from the gamma-ray and 
neutrino upper limits for $n=2.3$. In this case the 99.73\% CL region of the fit is in tension with the neutrino upper limit and marginally 
with the gamma-ray upper limit. Also in this case the gamma-ray upper limit loss restrictive power due to the attenuation in the EBL, in 
such a way that, the neutrino upper limit is more efficient to reject all models whose points fall in the 99.73\% CL region 
of the fit.

\begin{figure}[!ht]
\includegraphics[width=8cm]{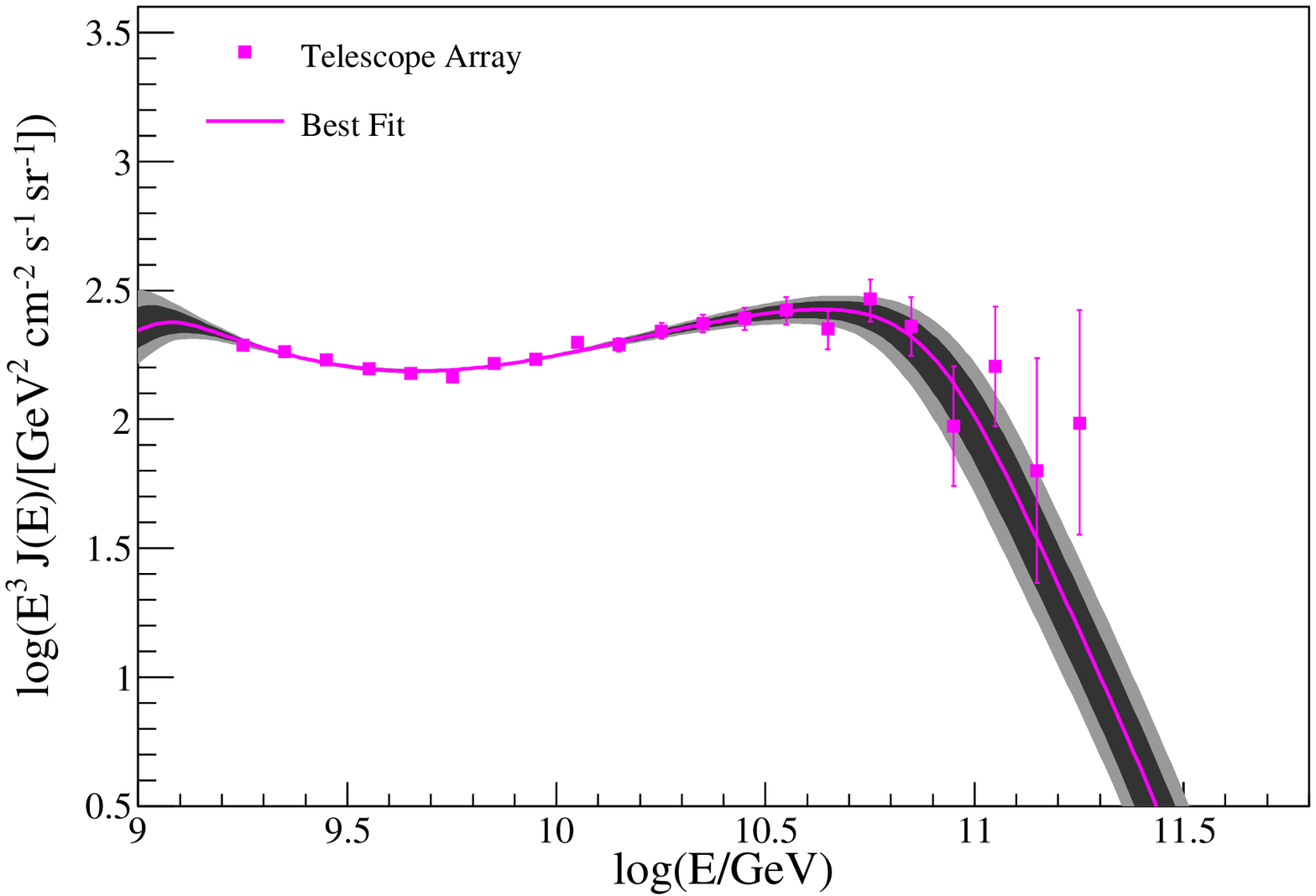}
\includegraphics[width=8cm]{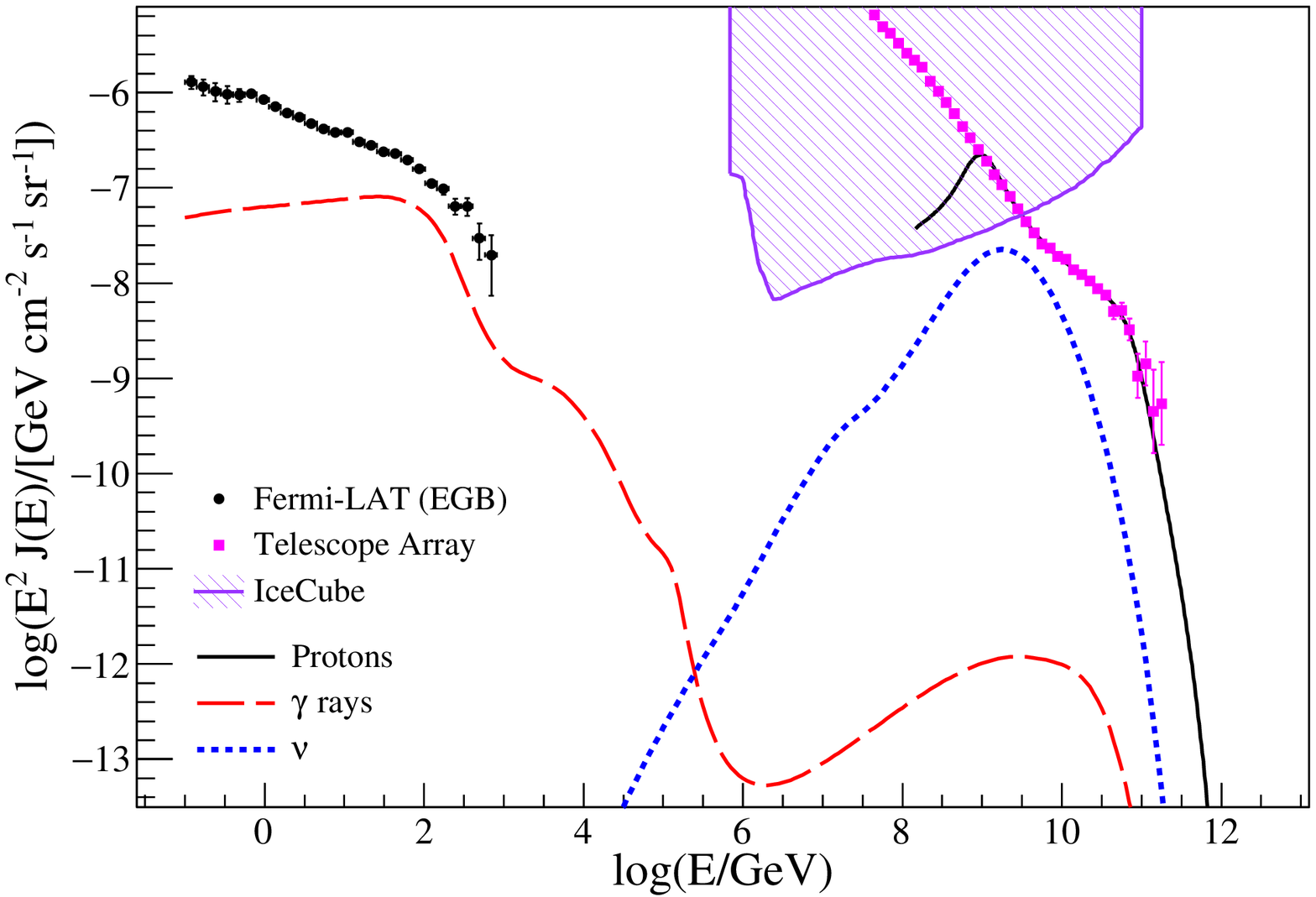}
\caption{Same as Fig.~\ref{KN21Inf} but for $E_{cut}=10^{19.7}$ eV. \label{KN197Inf}}
\end{figure}

Although $E_{cut} = 10^{21}$ eV is compatible with the Telescope array data \cite{Heinze:15}, the best fit value obtained in 
Ref.~\cite{Heinze:15} is $E_{cut} = 10^{19.7}$ eV with $\delta_E=-0.35$. Therefore, the same analysis done for $E_{cut} = 10^{21}$ eV 
is performed for this new value of the cutoff energy. Figure \ref{KN197Inf} shows the results obtained by using the EBL model of 
Ref.~\cite{Kneiske:04} and for no UHECR emission beyond $z=1$. The best fit parameters obtained in this case are $\alpha = 1.39$, 
$m = 8.34$, and $\delta_E = -0.31$. The values obtained in Ref.~\cite{Heinze:15} are $\alpha = 1.52$, $m = 7.7$, and $\delta_E = -0.35$ 
which are close to the ones obtained in this work. In any case, the differences can be explained by the different EBL model and 
propagation code used for the two calculations. Comparing the bottom panel of Fig.~\ref{KN197Inf} with the ones corresponding to 
Figs.~\ref{KN21Inf} and \ref{IN21Inf} it can be seen that for $E_{cut} = 10^{19.7}$ eV the production of gamma rays and neutrinos 
is smaller than for the case of $E_{cut} = 10^{21}$ eV. This is due to the fact that $E_{cut} = 10^{19.7}$ eV is very close to the 
threshold energy of photopion production undergone by protons interacting with the CMB photons. Therefore, this channel for the 
generation of these secondary particles is quite suppressed.         

\begin{figure}[!ht]
\includegraphics[width=8cm]{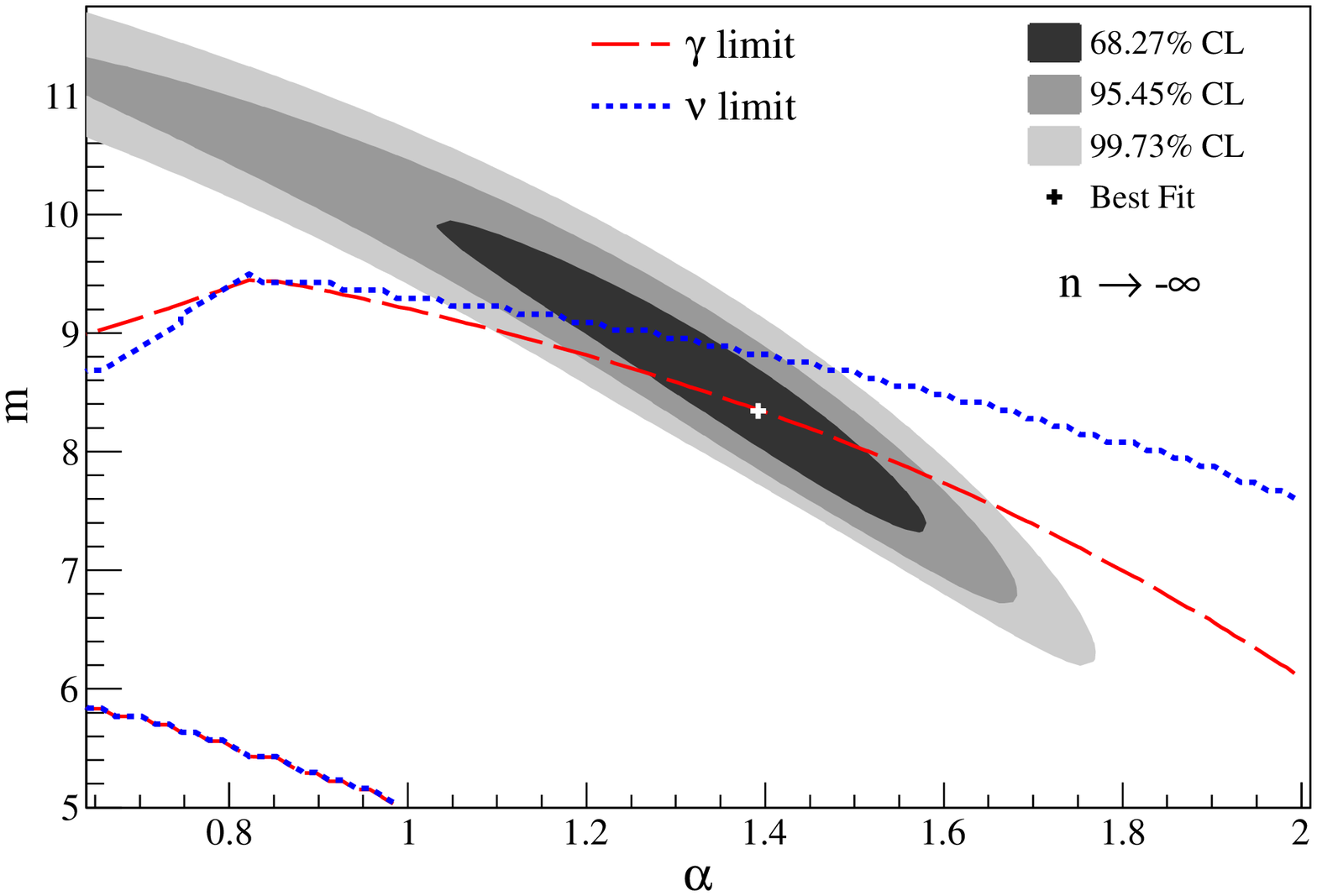}
\includegraphics[width=8cm]{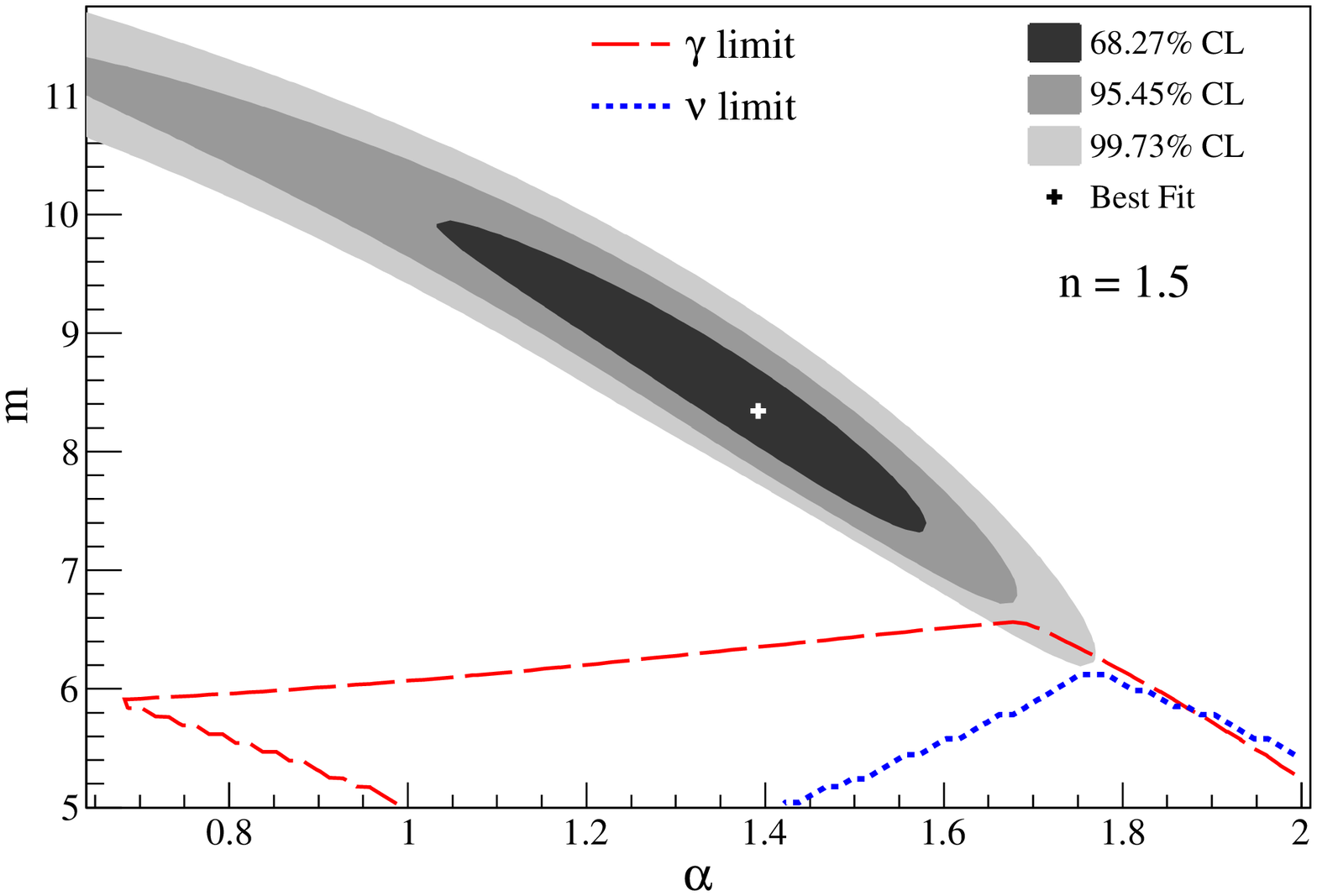}
\caption{Top panel: Best fit and confidence regions for $n \rightarrow -\infty$. The allowed regions corresponding to gamma-rays and neutrino 
observations are below the larger dashed and dotted curves and above the smaller ones on the left-bottom corner, respectively. Bottom panel: 
Best fit and confidence regions for $n=1.5$. The allowed region from the gamma-ray upper limit is below the large dashed curve and above the 
small dashed curve in the bottom-left corner. The allowed region from the neutrino upper limit is below the dotted curve. The EBL model of 
Ref.~\cite{Kneiske:04} is considered and $E_{cut}=10^{19.7}$ eV. \label{RegKN197}}
\end{figure}

The top panel of Fig.~\ref{RegKN197} shows the best fit parameters and the regions of 68.27\%, 95.45\%, and 99.73\% CL for the 
fit of Fig.~\ref{KN197Inf}. Also shown are the allowed regions inferred from the gamma-ray and neutrino upper limits. In this 
case, the best fit is compatible with the neutrino upper limit and it is marginally compatible with the gamma-ray upper limits. 
As expected, in this case the gamma-ray and neutrino upper limits are less restrictive than for $E_{cut} = 10^{21}$ eV. It is 
worth mentioning that the constraints imposed by the neutrino upper limit are less affected than the ones corresponding to the 
gamma rays, by the decrease of the cutoff energy. This is due to the fact that, as can be seen from the bottom panel of 
Fig.~\ref{KN197Inf}, the neutrino energy distribution becomes narrower than for the $E_{cut} = 10^{21}$ eV case and then the 
neutrino flux at the peak does not decrease considerably. The bottom panel of Fig.~\ref{RegKN197} shows the results corresponding 
to $n=1.5$ in which it can be seen that, for this value, the region of 99.73\% CL is in tension with the neutrino upper limit. 
In this case the neutrino upper limit is more restrictive than the one corresponding to the gamma-ray observations. 

Also for $E_{cut}=10^{19.7}$ eV a similar behavior is obtained when the EBL model of Ref.~\cite{Inoue:12} is considered. In this case, the best 
fit parameters are $\alpha = 1.38$, $m = 8.41$, and $\delta_E = -0.31$. For no UHECR emission above $z=1$ the best fit parameters fall in the
allowed regions inferred from the gamma-ray and neutrino upper limits and the 99.73\% CL region of the fit is in tension with the neutrino upper
limit for $n \gtrsim 1.6$.

\section{Conclusions}

In this work we have studied the constraints, imposed by the results recently obtained by the Fermi-LAT collaboration about the 
EGB origin, on proton models of UHECRs. For that purpose, we have first calculated an upper limit to the component of the 
integrated EGB flux, from 50 GeV to 2 TeV, that does not originate in point sources. The obtained value at 90\% CL is 
$I_\gamma^{UL} = 9.354 \times 10^{-10}\: \textrm{cm}^{-2} \textrm{s}^{-1} \textrm{sr}^{-1}$. 

We have assumed that the UHECR emission as a function of redshift $z$ follows a broken power law in $(1+z)$, with a breaking 
point at $z=1$ and an end point at $z=6$. We have found, by fitting the TA data, a very fast increase of the UHECR emission for 
$z \in [0,1]$, such that the index of the power law takes the values $m=6.65-8.34$, depending on the cutoff energy and EBL model 
considered. This result is consistent with previous work. We have also found that for $E_{cut} = 10^{21}$ eV the best fit, 
corresponding to the case in which there is no UHECR emission beyond $z=1$, is rejected at 99\% CL by using the inferred upper 
limit from gamma-ray observations. However, part of the 68.27\% CL region of the fit is in the allowed region inferred from that
upper limit. For $E_{cut} = 10^{19.7}$ eV the best fit is in the allowed regions obtained from the upper limits inferred from 
gamma-ray and neutrino observations.      

When the UHECR emission beyond $z=1$ is included the gamma-ray information becomes less restrictive. This is due to the larger
attenuation of the gamma-ray flux originated at $z>1$ in the energy interval considered for the integration. The process 
responsible for this attenuation is pair production on the EBL. In this case, the gamma-ray and neutrino observations become
complementary. By using both types of observations we have found that the index of the evolution function for $z \in [1,6]$ 
should be smaller than $n=1.5-2.3$ in order not to be in tension with the 99.73\% CL region of the fit. Therefore, only proton 
models of the UHECRs with a much slower redshift evolution in the interval $z \in [1,6]$, compared with the the one corresponding 
to $z \in [0,1]$, are still compatible with the neutrino and gamma-ray upper limits. A more precise determination of the 
different EGB components or a more restrictive neutrino upper limits are required to reject, at a given significance level, 
all proton models of UHECRs by using these types of analyses, which are independent of composition measurements.                
        
{\bf Note added:} as our paper was completed another manuscript appeared on the arXiv, considering the impact of the EGB 
measurements on proton models of UHECR, which contains complementary results to ours \cite{BereKala:16}.

\begin{acknowledgments}

A.~D.~S.~is a  member of the Carrera del Investigador Cient\'ifico of CONICET, Argentina. This work is supported by CONICET PIP 
2011/360, Argentina. The author thanks to O. Kalashev and E. Kido for their help with the TransportCR program and the members of the 
Pierre Auger Collaboration for useful discussions.  

\end{acknowledgments}

\appendix

\section{Upper limit on the integrated EGB flux that do not originate in point sources}
\label{UL}

\begin{figure}[!ht]
\includegraphics[width=8cm]{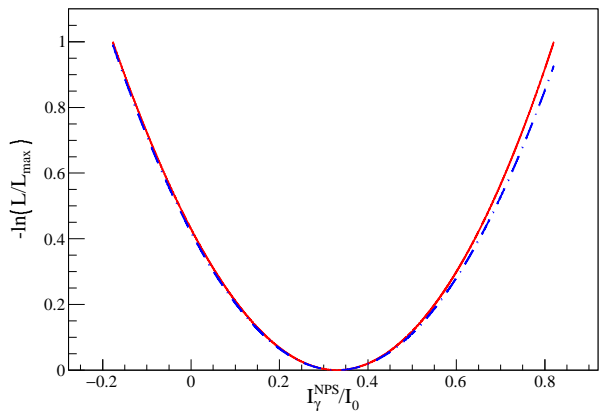}
\caption{$-\ln ( \mathcal{L}/\mathcal{L}_{max} )$ as a function of $\hat{I}_\gamma^{NPS}/I_0$, with $I_0=1\times 10^{-9}$ cm$^{-2}$s$^{-1}$sr$^{-1}$.
The solid line corresponds to the profile likelihood and the dotted-dashed line to its Gaussian approximation. \label{Lik}}
\end{figure}

In this appendix the calculation of the upper limit on the EGB flux component that does not originate in point sources, 
integrated from 50 GeV to 2 TeV, is presented. In Ref.~\cite{AckermannPS:16} it is shown that the integrated EGB flux, 
corresponding to the energy range from 50 GeV to 2 TeV and originated in point sources is 
$I_\gamma^{PS}=2.07^{+0.40}_{-0.34}\times 10^{-9}$ cm$^{-2}$s$^{-1}$sr$^{-1}$. The total integrated EGB flux corresponding 
to the same energy range is calculated in Ref.~\cite{Bechtol:15} and it is given by 
$I_\gamma^{EGB} = (2.4\pm0.3)\times 10^{-9}$ cm$^{-2}$s$^{-1}$sr$^{-1}$. Because $I_\gamma^{PS}$ has asymmetric errors the 
method developed in Ref.~\cite{Barlow:04} is used to find an approximated likelihood for the estimators $\hat{I}_\gamma^{EGB}$
and $\hat{I}_\gamma^{PS}$, which is given by,
\begin{eqnarray}
\ln \mathcal{L}(\hat{I}_\gamma^{EGB},\hat{I}_\gamma^{PS}) &=& -\frac{1}{2} \left( \frac{\hat{I}_\gamma^{EGB}-
\bar{I}_\gamma^{EGB}}{\sigma_{EGB}} \right)^2 \nonumber \\
&& -\frac{1}{2} \left( \frac{\hat{I}_\gamma^{PS}-\bar{I}_\gamma^{PS}}{\sigma_{PS}(\hat{I}_\gamma^{PS})} \right)^2,
\label{Lall}
\end{eqnarray}
where $\bar{I}_\gamma^{EGB}=2.4 \times 10^{-9}$ cm$^{-2}$s$^{-1}$sr$^{-1}$, 
$\sigma_{EGB}=0.3 \times 10^{-9}$ cm$^{-2}$s$^{-1}$sr$^{-1}$, and $\bar{I}_\gamma^{PS}=2.07 \times 10^{-9}$ cm$^{-2}$s$^{-1}$sr$^{-1}$.
Here, 
\begin{equation}
\sigma_{PS}(\hat{I}_\gamma^{PS})= \frac{ 2 \sigma_{+} \sigma_{-} }{ \sigma_{+} + \sigma_{-}}  + \frac{\sigma_{+} - \sigma_{-}}{\sigma_{+} + \sigma_{-}} %
\ (\bar{I}_\gamma^{PS}-\hat{I}_\gamma^{PS}),  
\end{equation}
where $\sigma_{+} = 0.40 \times 10^{-9}$ cm$^{-2}$s$^{-1}$sr$^{-1}$ and 
$\sigma_{-} = 0.34 \times 10^{-9}$ cm$^{-2}$s$^{-1}$sr$^{-1}$. The integrated EGB flux that do not originate in point 
sources is obtained by subtracting the contribution of the point sources to the total one, i.e. 
$I_\gamma^{NPS}=I_\gamma^{EGB}-I_\gamma^{PS}$. Therefore, introducing 
$\hat{I}_\gamma^{EGB}=\hat{I}_\gamma^{PS}+\hat{I}_\gamma^{NPS}$ in Eq.~(\ref{Lall}) and applying the profile likelihood 
method \cite{Agashe:14} to get rid of the nuisance parameter $\hat{I}_\gamma^{PS}$ an approximated likelihood function 
for $\hat{I}_\gamma^{NPS}$ is obtained. As expected, the minimum of $-\ln\mathcal{L}(\hat{I}_\gamma^{NPS})$ is attained 
at $0.33 \times 10^{-9}$ cm$^{-2}$s$^{-1}$sr$^{-1}$. Figure \ref{Lik} shows $-\ln ( \mathcal{L}/\mathcal{L}_{max} )$ as 
a function of $\hat{I}_\gamma^{NPS}/I_0$, where $\mathcal{L}_{max}$ is the value of the likelihood at the maximum and 
$I_0=1\times 10^{-9}$ cm$^{-2}$s$^{-1}$sr$^{-1}$. As can be seen from that figure the likelihood obtained is quite 
symmetric, this is due to the fact that the difference between $\sigma_{+}$ and $\sigma_{-}$ is not too big and that the 
likelihood corresponding to the point sources is combined with the one corresponding to the total EGB integrated flux, 
which is symmetric. In order to simplify the calculations a Gaussian approximation is developed where the mean value is 
given by $\tilde{I}_\gamma = 0.33 \times 10^{-9}$cm$^{-2}$s$^{-1}$sr$^{-1}$ and 
$\tilde{\sigma} = 0.36 \times 10^{-9}$ cm$^{-2}$s$^{-1}$sr$^{-1}$. The Gaussian approximation of the profile likelihood 
is also shown in the figure, it differs in less than $8\%$ in the two sigma region. In any case, it is chosen such that 
the Gaussian approximation is always smaller than the profile likelihood, which means that the upper limits obtained 
below are conservative. 

\begin{figure}[!hb]
\includegraphics[width=8cm]{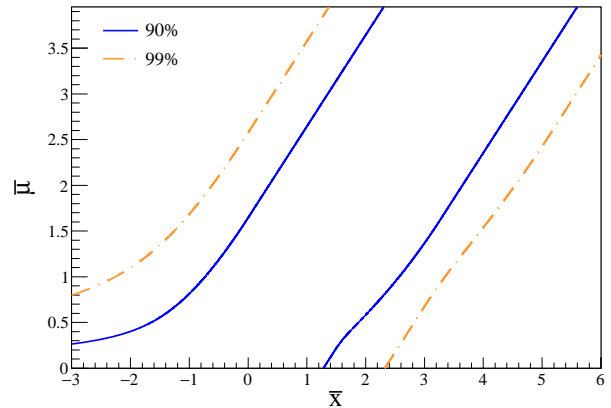}
\caption{The Feldman \& Cousins confidence belts for 90\% CL (solid lines) and 99\% CL (dotted-dashed lines) for a Gaussian distribution 
with unknown mean and standard distribution equal to 1. \label{FC}}
\end{figure}

The Feldman \& Cousins \cite{Feldman:98} method is applied in order to decide whether an upper limit or an interval containing the true value 
of $I_\gamma^{NPS}$, corresponding to a given CL, is obtained. By considering the following variable,
\begin{equation}
\bar{x}= \frac{I_\gamma^{NPS}}{\tilde{\sigma}},
\end{equation} 
the problem is reduced to the one corresponding to a Gaussian distribution with unknown mean, $\bar{\mu}$, and standard deviation equal to 1. 
Figure \ref{FC} shows the confidence belts for 90\% and 99\% CL. 

In this case,
\begin{equation}
\bar{x}= \frac{\tilde{I}_\gamma}{\tilde{\sigma}} = 0.93.
\end{equation} 
From Fig.~\ref{FC} it can be seen that, for this value, an upper limit is obtained for both, 90\% and 99\% CL. The results are,
\begin{equation}
I_\gamma^{UL} = \left\{ 
\begin{array}{ll}
9.345 \times 10^{-10}\ \textrm{cm}^{-2} \textrm{s}^{-1} \textrm{sr}^{-1} & 90\%\ \textrm{CL} \\
& \\
1.258 \times 10^{-9}\ \textrm{cm}^{-2} \textrm{s}^{-1} \textrm{sr}^{-1} & 99\%\ \textrm{CL}
\end{array}    \right.. 
\end{equation}

\section{Rejection regions inferred from the gamma-ray upper limit at 99\% CL}
\label{99}

\begin{figure}[!ht]
\includegraphics[width=8cm]{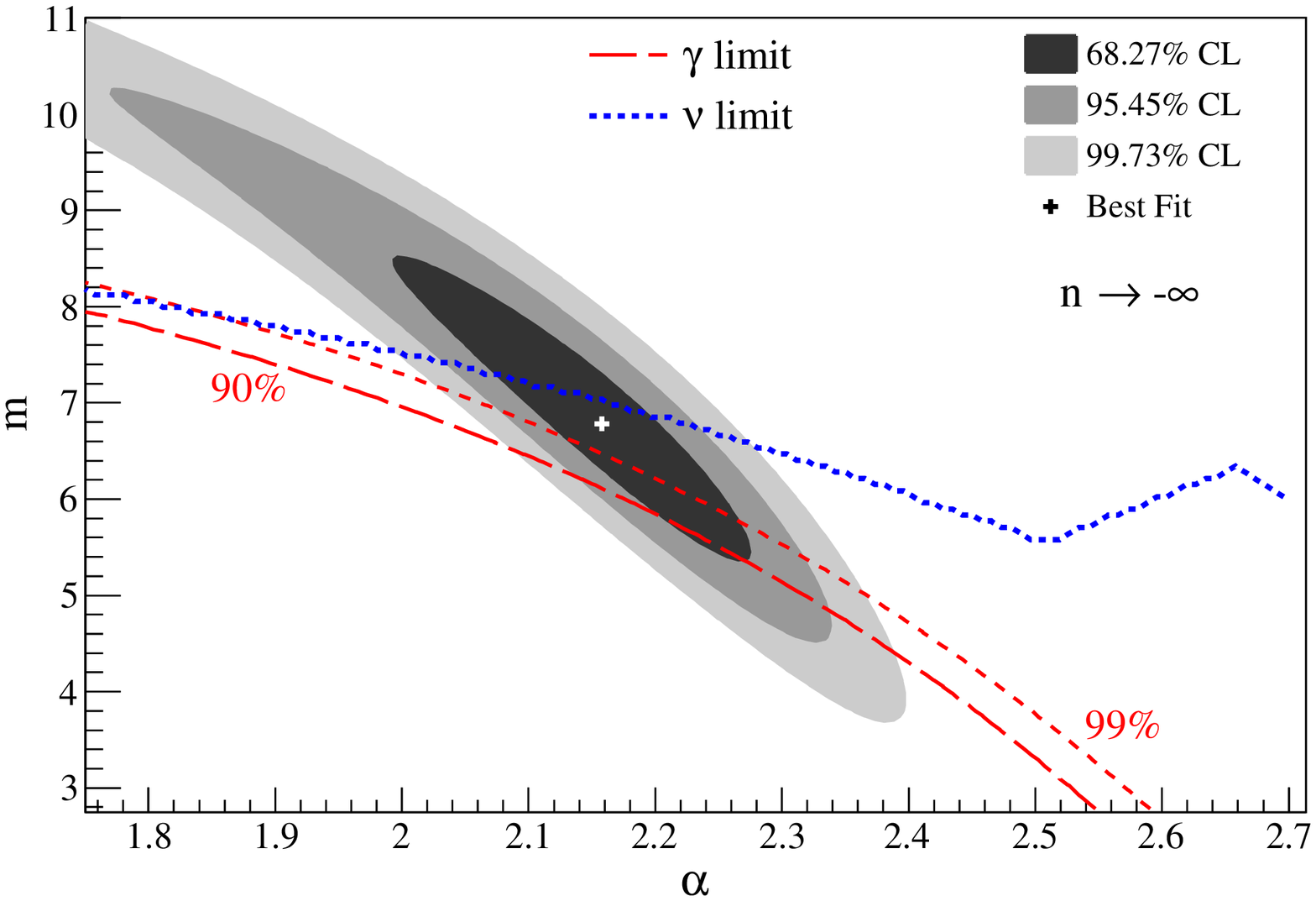}
\includegraphics[width=8cm]{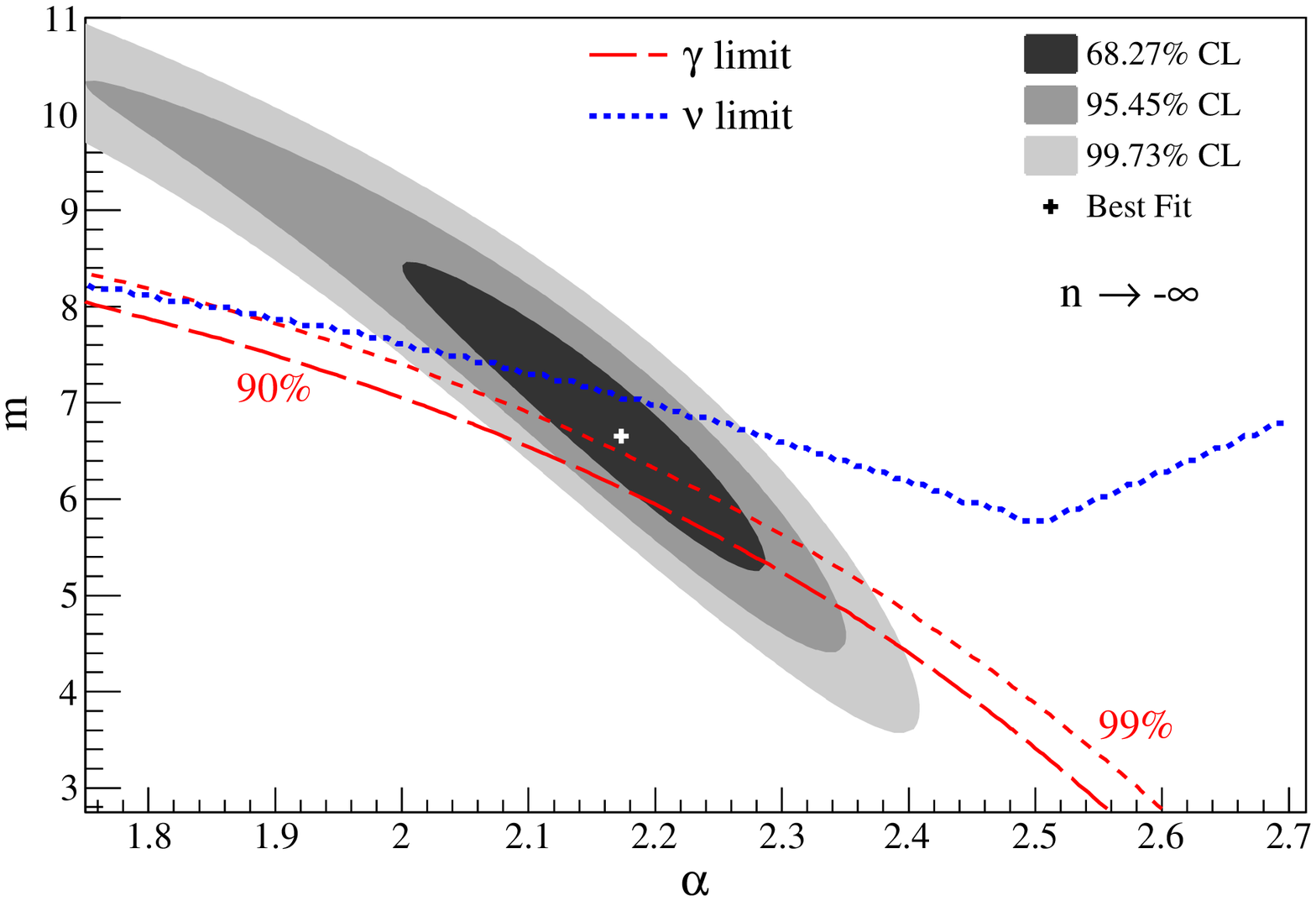}
\caption{Best fit and confidence regions for $n \rightarrow -\infty$. The allowed regions corresponding to gamma-rays 
and neutrino observations are below the dashed and dotted curves, respectively. The curves corresponding to the 
gamma-ray data are labelled with the rejection probability. The cutoff energy is $E_{cut}=10^{21}$ eV. The EBL models 
considered are the ones in Ref.~\cite{Kneiske:04} (top panel) and in Ref.~\cite{Inoue:12} (bottom panel). 
\label{Reg2199}}
\end{figure}

Figure \ref{Reg2199} shows the rejection regions obtained from the upper limit on the integrated EGB flux, corresponding 
to the energy range from 50 GeV to 2 TeV, that do not originate in point sources at 90 and 99\% CL (see Appendix \ref{UL}). 
The cases considered correspond to $E_{cut}=10^{21}$ eV and no UHECR emission beyond $z=1$. From the plot it can be seen 
that for both EBL models considered the best fit of the UHECR flux is rejected at 99\% CL. Also shown are the rejection 
regions inferred from the neutrino upper limit. 

\end{document}